\documentclass[runningheads,a4paper]{llncs}

\usepackage{amssymb}
\usepackage{graphicx}

\usepackage{url}

\usepackage{makeidx}  % allows for indexgeneration
\usepackage{graphicx}

 \usepackage{eurosym}
\usepackage{multirow}
\usepackage{float}
\usepackage{environ}
\usepackage{datetime}
\usepackage{mathtools}
\usepackage{alltt}
\usepackage{array}
\newcolumntype{P}[1]{>{\centering\arraybackslash}p{#1}}
\usepackage{color}
\usepackage[ruled,vlined]{algorithm2e}
\usepackage{eurosym}
\usepackage{amsmath}
\usepackage{wasysym}
\usepackage{silence}
\makeatletter
\newlength{\phaserulewidth}
\newcommand{\setphaserulewidth}{\setlength{\phaserulewidth}}

\makeatother

\setphaserulewidth{.7pt}

\usepackage[space]{grffile}
\usepackage{array}
\usepackage[table]{xcolor}
\usepackage{wasysym}
\usepackage{pbox}

\usepackage{eurosym}
\usepackage{multirow}
\DeclareGraphicsExtensions{.jpg,.png}
\urldef{\mailsa}\path|{alfred.hofmann, ursula.barth, ingrid.haas, frank.holzwarth,|
\urldef{\mailsb}\path|anna.kramer, leonie.kunz, christine.reiss, nicole.sator,|
\urldef{\mailsc}\path|erika.siebert-cole, peter.strasser, lncs}@springer.com|    
\newcommand{\keywords}[1]{\par\addvspace\baselineskip
\noindent\keywordname\enspace\ignorespaces#1}

\begin{document}

\mainmatter  % start of an individual contribution

\title{Performance Evaluation of a Distributed Clustering Approach for Spatial Datasets}

\titlerunning{PE of a Distributed Clustering Approach for Spatial Datasets.}

\author{Malika Bendechache\inst{1} \and Nhien-An LE-KHAC\inst{2} \and M-Tahar Kechadi\inst{1}}
\authorrunning{Malika Bendechache et al.} % abbreviated author list (for running head)
\tocauthor{Malika Bendechache,  Nhien-An LE-KHAC and M-Tahar Kechadi}
\authorrunning{Malika Bendechache et al.}

\institute{Insight centre for Data Analytics \\
University College Dublin, \\
O'Brien building, Centre East\\
Belfield, Dublin 04, \\
\email{malika.bendechache@gamil.com},\\
\email{tahar.kechadi@ucd.ie},\\
\and
University College Dublin \\
Belfield, Dublin 04\\
\email{an.lekhac@ucd.ie}}

\toctitle{Performance Evaluation of a Distributed Clustering Approach for Spatial Datasets}
\tocauthor{Malika Bendechache et al.}
\maketitle

\begin{abstract}
The  analysis of  big  data requires  powerful, scalable,  and  accurate data  analytics
  techniques that the traditional data mining and machine learning do not have as a whole.
  Therefore, new data analytics frameworks are needed to deal with the big data challenges
  such  as volumes,  velocity, veracity,  variety of  the data.   Distributed data  mining
  constitutes a  promising approach  for big data  sets, as they  are usually  produced in
  distributed  locations,   and  processing  them   on  their  local  sites   will  reduce
  significantly the  response times, communications,  etc.  In  this paper, we  propose to
  study the performance of a distributed clustering, called Dynamic Distributed Clustering
  (DDC). DDC has the  ability to remotely generate clusters and  then aggregate them using
  an efficient aggregation algorithm.  The technique is developed for spatial datasets. We
  evaluated the DDC using two types  of communications (synchronous and asynchronous), and
  tested using various load distributions. The experimental results show that the approach
  has super-linear  speed-up, scales up  very well, and can  take advantage of  the recent
  programming models,  such as  MapReduce model, as  its results are  not affected  by the
  types of communications.
\end{abstract}

\section{Introduction}
\label{sec:Int}

Nowadays big data is becoming a commonplace.  It is generated by multiple sources at rapid
pace, which leads to very large data volumes that need to be stored, managed, and analysed
for useful insights. From organisations point of view, it is not the size of the generated
data which  is important.   It is what  we learn from  it that  matters, as this  may help
understanding the behaviour  of the system that is  governed by this data or  help to make
some key decisions,  etc. To extract meaningful  value from big data,  we need appropriate
and efficient mining and analytics techniques to analyse it.  One of the most powerful and
common   approaches   of  analysing   datasets   for   extracting  useful   knowledge   is
clustering. Clustering has a wide range of  applications and its concept is so interesting
that numerous  algorithms for various  types of data  have been proposed  and implemented.
However, big data come  up with new challenges, such as  large volumes, velocity, variety,
and veracity, that  the majority of popular clustering algorithms  are inefficient at very
large scale. This inefficiency  can be that the final results are  not satisfactory or the
algorithm has  high complexity which requires  large computing power and  response time to
produce  final  results.  There  are  two  major categories  of  approaches  to deal  with
computational complexity of these clustering algorithms: 1) the first category consists of
reducing the size of  the initial dataset.  One can use  either sample-based techniques or
dimensionality reduction  techniques. The second  category consists of using  parallel and
distributed  computing to  speed  up  the response  time.   In this  case  we  can try  to
parallelise or model  the algorithm in the  form of a client-server  model using MapReduce
mechanism.   However,  these  algorithms  are inherently  difficult  to  parallelise,  and
designing  an  efficient distributed  version  of  the  algorithm is  not  straightforward
either. This  is due  to the fact  that the  processing nodes, either  in the  parallel or
distributed versions, need to communicate and  coordinate their efforts in order to obtain
the same results.  These communications are extremely expensive and can cancel the benefit
of the parallelised version.  To deal with these challenging issues, we propose to study a
distributed approach  that takes  advantage of parallel  and distributed  computing power,
while getting ride of the drawbacks of the  previous methods. In addition, one of the main
advantages of  our approach  is that  it can  be used  as a  framework for  all clustering
algorithms.  In other words, while it is  well known that there is no clustering algorithm
that can universally be used to cluster every dataset of a given application, our approach
can be used for  all algorithms or a set of algorithms to  derive a distributed clustering
approach for a given data having specific characteristics.

The proposed  approach has two  main phases:  the first phase,  based on the  SPMD (Single
Program Multiple  Data) paradigm, consists of  dividing the datasets into  $K$ partitions,
where $K$ is the number of processing nodes.  Then, for each partition we cluster its data
into $C_i$  clusters. This phase  is purely parallel, as  each processing node  executes a
clustering  algorithm on  its data  partition independently  of the  others. The  obtained
clusters  on each  node  are called  local  clusters.   This phase  does  not require  any
communications, and in addition, in the majority  of applications the data is collected by
various sources,  which are  geographically distributed.  Therefore,  the data  is already
partitioned.  All required is  to cluster locally the data.  The  second phase consists of
aggregating  (or  merging)  the  local  clusters to  obtain  global  clusters  by  merging
overlapping clusters.  In  order to determine whether two local  clusters belonging to two
different nodes are overlapping  or not, one needs to exchange  the local clusters between
the nodes.   This operation is  extremely expensive when the  dataset is very  large.  The
main idea of our approach is to minimise the data exchange while maximising the quality of
the global  clusters.  The  method used  to aggregate spatial  local clusters  into global
clusters allows only  to exchange about $2\%$ of the  original datasets ~\cite{Laloux-11},
which  is highly  efficient.  In  this paper,  we want  to study  the performance  of such
distributed clustering  technique by  calculating its speedup  compared to  the sequential
version of the algorithm, its scalability, its communication overheads, and its complexity
in general.

The  rest of  the paper  is organised  as follows:  In the  next section  we will  give an
overview of  the state-of-the-art of parallel  and distributed data mining  techniques and
discuss their limitations.  Then we will  present in more details the proposed distributed
framework  and its  concepts in  Section \ref{sec:DDC}.   In section  \ref{sec:DDC-EV}, we
evaluate  its  performance  based  on   two  types  of  implementations;  synchronous  and
asynchronous  communications.  In  section  \ref{sec:ExRes}, we  discuss the  experimental
results  based on  speedup,  scalability,  communication overheads,  and  compare the  two
implementation  models; synchronous  and  asynchronous. Finally,  we  conclude in  Section
\ref{sec:Con}.

\section{Related Work}
\label{Sec:RW}

Distributed Data Mining  (DDM) is a line  of research that has attracted  much interest in
recent years~\cite{han2011data}.  DDM  was developed because of the need  to process data that
can  be very  large or  geographically distributed  across multiple  sites.  This  has two
advantages: first,  a distributed system has  enough processing power to  analyse the data
within a reasonable time  frame. Second, it would be very advantageous  to process data on
their respective sites to avoid the transfer of  large volumes of data between the site to
avoid heavy communications, network bottlenecks, etc. 

DDM techniques can be  divided into two categories based on  the targeted architectures of
computing  platforms~\cite{Zaki-00}. The  first,  based on  parallelism, uses  traditional
dedicated and  parallel machines with  tools for communications between  processors. These
machines are generally  called super-computers.  The second category targets  a network of
autonomous  machines.  These  are called  distributed  systems, and  are characterised  by
distributed  resources, low-speed  network  connecting the  system  nodes, and  autonomous
processing  nodes which  can be  of different  architectures, but  they are  very abundant
~\cite{ghosh-14}.  The main goal of this category  of techniques is to distribute the work
among the  system nodes and try  to minimise the  response time of the  whole application.
Some   of   these   techniques   have   already  been   developed   and   implemented   in
~\cite{Aouad-07s,Wu-14}.

However, the  traditional DDM methods  are not always effective,  as they suffer  from the
problem of scaling. One solution to deal with  large scale data is to use parallelism, but
this is very  expensive in terms of communications and  processing power. Another solution
is to reduce the size of training  sets (sampling).  Each system node generates a separate
sample.    These   samples   will   be   analysed  using   a   single   global   algorithm
~\cite{Zhang-96,Jain-99}.  However, this technique has a disadvantage that the sampling in
this case  is very complex  and requires many communications  between the nodes  which may
impact on the quality of the samples and therefore the final results.  This has led to the
development of techniques that rely on ensemble learning ~\cite{Rokach-14,Bauer-99}. These
new techniques are very  promising, as each technique of the  ensemble network attempts to
learn from the data and the best or compromised results of the  network will emerge as the
winner. Integrating ensemble learning methods in DDM framework will allow to deal with the
scalability problem, as it is the case of the proposed approach. 

Clustering algorithms  can be divided  into two  main categories, namely  partitioning and
hierarchical. Different elaborated taxonomies of  existing clustering algorithms are given
in the literature.  Many parallel clustering  versions based on these algorithms have been
proposed  in  the  literature~\cite{aouad-07lightweight,Dhillon-99,Ester-96,Garg-06,Geng-05,Inderjit-00,Xu-99}.  These  algorithms are further classified  into two sub-categories.
The first consists of methods requiring  multiple rounds of message passing.  They require
a  significant amount  of synchronisations  and  data exchange.   The second  sub-category
consists of methods that build local clustering models  and send them to a central site to
build global models ~\cite{Laloux-11}.

In ~\cite{Dhillon-99} and ~\cite{Inderjit-00}, message-passing versions of the widely used
K-Means algorithm were proposed.  In ~\cite{Ester-96} and ~\cite{Xu-99}, the authors dealt
with   the   parallelisation   of   DBSCAN;  density-based   clustering   algorithm.    In
~\cite{Garg-06} a parallel  message passing version of the BIRCH  algorithm was presented.
A parallel version of a hierarchical  clustering algorithm, called MPC for Message Passing
Clustering,  which  is   especially  dedicated  to  Microarray  data   was  introduced  in
~\cite{Geng-05}.  Most of  the parallel approaches need  either multiple synchronisation
constraints   between   processes  or   a   global   view   of   the  dataset,   or   both
~\cite{aouad-07lightweight}. All these approaches deal  with the parallelisation of the sequential
version  of the  algorithm by  trying phases  of the  algorithm which  can be  executed in
parallel by  several processors.  However,  this requires many synchronisations  either to
access shared  data (for the shared  memory model) or communications  (for message passing
model).   In  some  algorithm  these synchronisations  and  communications  are  extremely
expensive and it is not worth parallelising them. This approach is not usually scalable.

In~\cite{brecheisen-06}  a client-server  model  is  adopted, where  the  data is  equally
partitioned and distributed among the servers, each of which computes the clusters locally
and  sends back  the results  to the  master. The  master merges  the partially  clustered
results to obtain  the final results.  This strategy incurs  a high communication overhead
between the master and  slaves, and a low parallel efficiency  during the merging process.
Other    parallelisations     using    a     similar    client-server     model    include
~\cite{arlia-01,chen-10,coppola-02,fu-11,guo-02,zhou-00}.  Among these approaches, various
programming  mechanisms  have been  used,  for  example,  a special  parallel  programming
environment, called  skeleton based programming in~\cite{coppola-02}  and parallel virtual
machine in~\cite{guo-02}. A Hadoop-based approach is presented in~\cite{fu-11}.

Another approach presented  in ~\cite{aouad-07lightweight} also applied a merging  of local models
to create the global models. Current approaches  only focus on either merging local models
or  mining a  set  of local  models  to build  global  ones. If  the  local models  cannot
effectively  represent local  datasets  then  global models  accuracy  will  be very  poor
~\cite{Laloux-11}.  In addition,  both partitioning and hierarchical  categories have some
issues which are very  difficult to deal with in parallel  versions.  For the partitioning
class, it needs  the number of clusters to  be fixed in advance, while in  the majority of
applications  the  number  of classes  is  not  known  in  advance. For  the  hierarchical
clustering  algorithms,  they  have  the  issue  of  stopping  conditions  for  clustering
decomposition, which is not an easy task and mainly in distributed versions.

\section{Dynamic Distributed Clustering}
\label{sec:DDC}
Dynamic Distributed Clustering  (DDC) model is introduced to deal  with the limitations of
the  parallel  and  master-slave  models.    DDC  combines  the  characteristics  of  both
partitioning and  hierarchical clustering methods.   In addition, it does  neither inherit
the problem of the number of partitions to be fixed in advance nor the problem of stopping
conditions. It is  calculated dynamically and generates global clusters  in a hierarchical
way. All these features look very promising  and some of them have been thoroughly studies
in ~\cite{Bendechache-16a}, such as the dynamic  calculation of the number of the clusters
and the accuracy  of the final clustering, in  this study one wants to show  the effect of
the communications on the response time,  the communication model used, the scalability of
the approach, and finally its performance in  terms of speed up compared to the sequential
version. In this paper we will focus on
\begin{itemize}
   \item Synchronous and asynchronous communications,  as this approach can be implemented
     either with synchronous or asynchronous communications. Both implementations produce 
     the same results. 
   \item  The speed-up  of  the DDC  approach  using  DBSCAN as  the  basic algorithm  for
     clustering the partitions. This algorithm  is known to have non-polynomial complexity
     ($O(n^2)$). 
   \item Scalability of the approach as the size of the dataset increases.
\end{itemize}

We start by briefly explaining the algorithm and then present a performance and evaluation
model for the approach.

\begin{figure}[!htb]
    \centering
    \begin{center}
    \includegraphics[width=\textwidth]{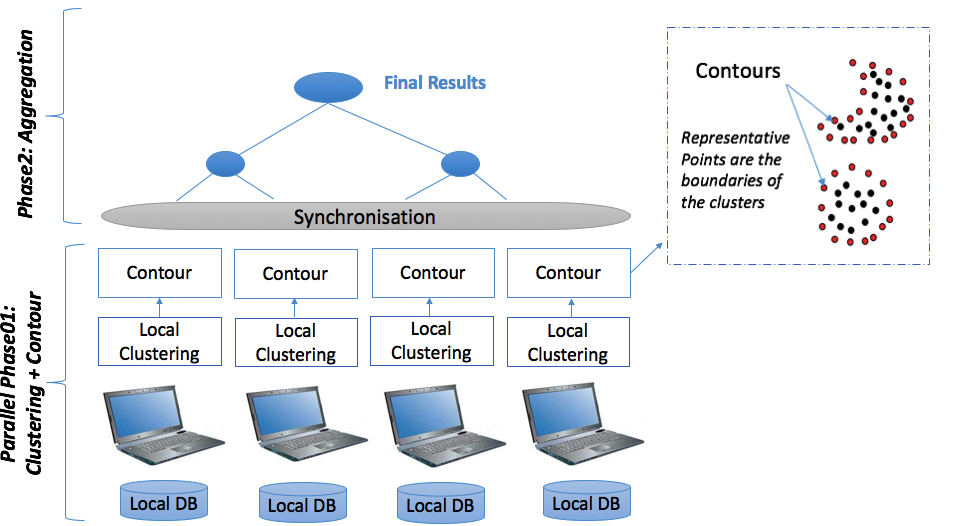}
    \caption{An overview of the DDC Approach.}
    \label{Archi}
     \end{center}
\end{figure}

The DDC approach has two main phases.  In the first phase, we cluster the datasets located
on  each processing  node and  select good  local representatives.   All local  clustering
algorithms are executed  in parallel without communications between the  nodes.  As DBSCAN
is the basic algorithm for clustering local datasets, we can reach a super linear speed-up
of $p^2$,  where $p$ is  the number  of processing nodes.   The second phase  collects the
local clusters from each node and affects them to some special nodes in the system; called
leaders.  The  leaders are elected  according to  their characteristics such  as capacity,
processing power,  connectivity, etc. The  leaders are  responsible for merging  the local
clusters.  In the following  we explain how the local clusters  are represented and merged
to generate global clusters.

\subsection{\textbf{Local Models}}

The local clusters are highly dependent on  the clustering techniques used locally in each
node. For instance,  for spatial datasets, the  shape of a cluster is  usually dictated by
the technique used to obtain them.  Moreover, this is not an issue for the first phase, as
the accuracy of  a cluster affects only the  local results of a given  node.  However, the
second phase  requires sending and  receiving all local clusters  to the leaders.   As the
whole data is  very large, this operation  will saturate very quickly the  network. So, we
must avoid  sending all the  original data through  the network. The  key idea of  the DDC
approach is to send only the cluster's representatives, which constitute between $1\%$ and
$2\%$  of the  whole  data.  The  cluster  representatives consist  of  the internal  data
representatives plus the boundary points of the cluster.
 
There are  many existing data  reduction techniques in the  literature.  Many of  them are
focusing only on the  dataset size. For instance, they try to  reduce the storage capacity
without paying  attention to the knowledge  contained in the data.   In ~\cite{N-A-10}, an
efficient reduction technique has been proposed;  it is based on density-based clustering.
Each  cluster  is  represented  by  a   set  of  carefully  selected  data-points,  called
representatives.   However, selecting  representatives is  still a  challenge in  terms of
quality and size ~\cite{Januzaj-04,Laloux-11}.
 
The best way  to represent a spatial cluster is  by its shape and density. The  shape of a
cluster   is  represented   by  its   boundary   points  (called   contour)  (see   Figure
\ref{Archi}). Many algorithms for extracting the boundaries from a cluster can be found in
the literature ~\cite{M.J-04,A.Ray-97,M.Melkemi-00,Edelsbrunner-83,A.Moreira-07}. We use
an   algorithm   based  on   triangulation    to   generate   the   clusters'   boundaries
~\cite{M.Duckhama-08}.   It   is  an  efficient  algorithm   for  constructing  non-convex
boundaries.  It is able to accurately characterise  the shape of a wide range of different
point distributions and densities with a reasonable complexity of $\mathcal{O}(n\log{n})$.

\subsection{\textbf{Global Models}}

The global  clusters are  generated in the  second phase  of the DDC.  This phase  is also
executed  in a  distributed fashion  but, unlike  the first  phase, it  has communications
overheads.  This  phase consists of two  main steps, which  can be repeated until  all the
global clusters  were generated.  First, each  leader collects the  local clusters  of its
neighbours.   Second,  the  leaders  will  merge the  local  clusters  using  the  overlay
technique. The process of merging clusters will continue until we reach the root node. The
root node will contain the global clusters (see Figure \ref{Archi}).

In  DDC we  only  exchange the  boundaries  of  the clusters.  The  communications can  be
synchronous   or  asynchronous.   We  implemented   this   phase  using   both  types   of
communications. An evaluation model is presented in the next Section.

The  pseudo  codes  of the  two  phases  of  the  DDC  framework are  given  in  Algorithm \ref{Algo-Phase1} and Algorithm \ref{Algo-Phase2}.

\begin{algorithm}[!htb]
 Initialisation\\
  $Node_i \in N$, $N$: The total nodes in the system.
    
  \SetKwInOut{Input}{input}\SetKwInOut{Output}{output}
  \Input {$X_i$: Dataset Fragment, $Params_i$: Input parameters for the local clustering:
    $Params_i$= ($Eps_i$, $MinPts_i$) for DBSCAN for example}
  \Output {$C_i$: Cluster's contours of $Node_i$ }
  \BlankLine
  \BlankLine
  
  \ForEach {$Node_i$ } {
    $L_i$  = Local\_Clustering($X_i$,$Params_i$)\;  \tcp {$Node_i$  executes a  clustering
      algorithm locally.}
    $C_i$ = Contour($L_i$)\; \tcp {$Node_i$ executes a contour algorithm locally.}
  } 
  
  \Return ($C_i$)\;
  \label{Algo-Phase1}
  \caption{DDC algorithm : Phase 1: Local clustering}
  
\end{algorithm}
%%%%%%%%%%%%%%%%%%%%%%%%%%%% 
\begin{algorithm}[!htb]

  \SetKwInOut{Input}{input}\SetKwInOut{Output}{output}
  \Input {$D$: Tree degree, $C_i$: Local
    cluster's contours generated by $Node_i$ in the phase 1} 
  \Output {$C_{G_{k,level}}$ : Global Cluster's contours (global results, level=0)}
  \BlankLine
  \BlankLine
  
  \Repeat{(level == 0)}{
    
    $level = treeHeight$\; 
    $Node_i$ joins a group $G_{K, {Level}}$  of $D$ elements\; \tcp{$Node_i$ joins its neighbourhood}
    $Node_j$=ElectLeaderNode($G_{K, {Level}}$)\;  \tcp{ $Node_j$ is the leader of the group $G$}
    
    \tcp{In parallel}
    \ForEach {$Node_i \in G_{K, {Level}}$} 
    {%
      
      \uIf {($i<>j$)}{%
        Send ($C_i$, $Node_j$)\;
        \tcp{Each node sends its contours to others nodes in the same group of neighbourhood}
      }
      \Else 
      {
        Recv ($C\equiv  (\{C_l\}$, $Node_l$))\; \tcp{If  the node  is the leader,  it will
          receive the others node's contours in the same group of neighbourhood}

        $G_{K,{Level}}$= Merge ($C_i$, $C$)\;\tcp{Merge the overlapping contours }
        
      }
    }
    $level$ - - \; 
  }
  \Return{($C_{G_{k,0}}$)}
  \label{Algo-Phase2}
  \caption{DDC algorithm : Phase 2: Merging}
  
\end{algorithm}

\section{DDC Evaluation}
\label{sec:DDC-EV}
In  order  to evaluate  the  performance  of the  DDC  approach,  we use  different  local
clustering algorithms. For  instance, with both K-Means  ~\cite{bendechache-15} and DBSCAN
~\cite{Bendechache-16a,Bendechache-16b}, the DDC  approach outperforms existing algorithms
in both quality of  its results and response time including K-Means  and DBSCAN applied to
the  whole dataset.   In this  section we  evaluate its  speed-up, scalability,  and which
architecture is  more appropriate to  implement it. In addition,  we compare DDC  with the
sequential version of the basic clustering algorithm used within DDC.

%\subsection{Synchronous communication}
%\label{synS}
The  proposed approach  is  more  developed for  distributed  systems  than pure  parallel
systems.   Therefore,  it  is  worth  analysing  the  benefits  of  using  synchronous  or
asynchronous  processing  mechanism,  as  distributed systems  are  asynchronous  and  the
blocking operations have a strong impact in communication time~\cite{solar-13}.

\begin{figure}[!htb]
\begin{center}
\includegraphics[width=\columnwidth]{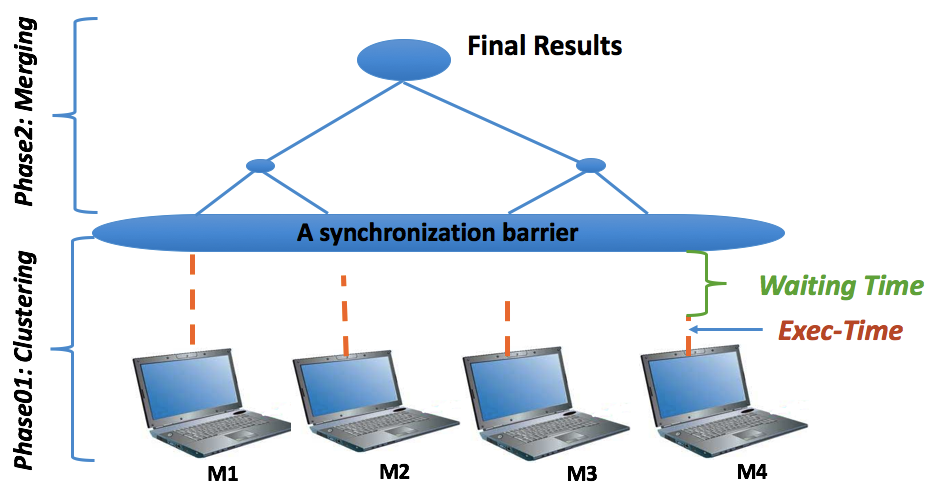}
\caption{Synchronous communications.}
\label{syn}
\end{center}
\end{figure}

In synchronous  model, as  illustrated in  Figure \ref{syn},  although machines  $M_3$ and
$M_4$ have  finished their computations  before $M_1$ and $M_2$,  they can not  send their
results until $M_1$  and $M_2$ finish as  well.  In this model, Not  only the computations
and communications are  not overlapped but also the machines  which finished early wasted
sometime waiting for the other to finish ~\cite{solar-13}.

\begin{figure}[!htb]
\begin{center}
\includegraphics[width=\columnwidth]{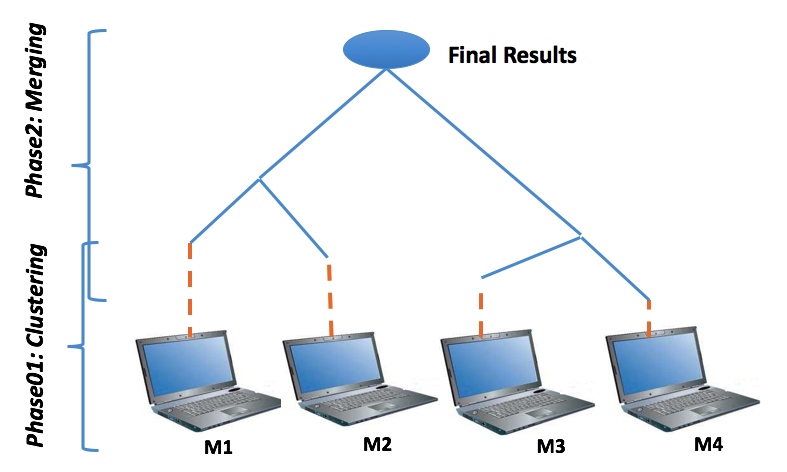}
\caption{Asynchronous communications.}
\label{asyn}
\end{center}
\end{figure}

In asynchronous model the machines which finished  early can advance to the next step. The
machines manage their communications and the 1st and 2nd phase overlap. This model is much
more  suitable for  distributed  computing,  where the  nodes  are  heterogeneous and  the
communications are usually slow. As can be seen in the example given in Figure \ref{asyn},
$M_3$  and  $M_4$  start  merging  their  results before  $M_1$  and  $M_2$  finish  their
computations.

\subsection{DDC Computational Complexity}
\label{CC}
Let $M$  be the number  of nodes and  $n_i$ the  dataset given to  each node $v_i$  in the
system.  The complexity  of our approach is  the sum of its  components' complexity; local
mining, local reduction, and global aggregation.

{\bf Phase1 - Local clustering:} Let $\Gamma(n_i)$ denote the local clustering
algorithm running on node  $(v_i)$, and $\Delta(c_i)$ be the time  required to execute the
reduction algorithm.  The cost of this phase is given by:
\begin{equation}
  \label{eq:1}
  T_{Phase_1} = \operatorname*{Max}_{i=1}^M(\Gamma(n_i) + \Delta(c_i)) 
\end{equation}

Where  $c_i$ is  the cluster  points  generated by  node  $v_i$. Note  that the  reduction
algorithm is of complexity $\mathcal{O}(c_i\log{}c_i)$. 

{\bf Phase2  - Aggregation:} The  aggregation depends  on the hierarchical  combination of
contours of local clusters. As the combination  is based on the intersection of edges from
the contours,  the complexity  of this  phase is  $\mathcal{O}(w_i\log{}w_i +  p)$.  Where
$w_i$ is  the total vertices  of the contours  by node $v_i$  and $p$ is  the intersection
points between edges of different contours (polygons).

{\bf Total complexity:} The total complexity of  the DDC approach, assuming that the local
clustering algorithm is DBSCAN which is of complexity $\mathcal{O}(n^2)$, is:
\begin{equation}
  \label{eq:2}
  T_{Total}  = \mathcal{O}(n_i^2)+\mathcal{O}(c_i\log{}c_i) + \mathcal{O}(w_i\log{}w_i+p)
\simeq \mathcal{O}(n_i^2)
\end{equation}

\subsection{DDC Speedup}
\label{SD}
The  DDC speedup  is  calculated against  the  sequential version  of  the approach.   The
sequential version  consists clustering all  the data on  one machine. Therefore,  it does
require  neither reduction  nor  aggregation.  Let  $T_1$  be the  execution  time of  the
sequential version  and $T_p$  the execution  time of the  DDC on  $p$ nodes.  The speedup
$\alpha$ is given by
\begin{equation}
  \label{eq:3}
    \alpha = \frac{T_1}{T_p}
\end{equation}

% \begin{figure}[!htb]
% \begin{center}
% \includegraphics[width=0.5\columnwidth]{speedup.png}
% \caption{Speedup.}
% \label{speed}
% \end{center}
% \end{figure}

Note that  if the complexity  of the clustering algorithm  is polynomial then  the optimal
speedup that can be  reached is $P$, under the condition that there  is no overhead due to
communications  and  extra  work.   If  the complexity  of  the  clustering  algorithm  is
$\mathcal{O}(n^2)$  then the  optimal speedup  can  be $P^2$;  this is  called {\em  super
  speedup}. In the following section we will evaluate the speedup in the case of DBSCAN.

\section{Experimental Results}
\label{sec:ExRes}

We  have implemented  our  approach on  a distributed  computing  system. The  distributed
computing system  consists of heterogeneous  desktops (different CPUs, OSs,  memory sizes,
loads, etc.). We use JADE (Java Agent DEvelopment), as a development platform to implement
the  approach.  JADE  is based  on  a 2P2  communication  architecture, It  allows to  use
heterogeneous processing  nodes, it  is scalable, and  dynamic ~\cite{cortese-05benchmark,bellifemine-05}.

The system nodes (desktops) are connected to local area networks. This allows us to add as
many  nodes as  required, depending  on the  experiment. Table  \ref{capa} lists  types of
machines  used to  perform the  experiments.  The  main goal  here is  to demonstrate  the
performance of the DDC in a heterogeneous distributed computing environment.

\begin{table}[!htb]
\centering
\caption{The characteristics of the used Machines}
\label{capa}
\begin{tabular}{|c|c|c|c|}
\hline
\textbf{Machine's name} & \textbf{Operating System}   & \textbf{Processor}   & \textbf{Memory}  \\\hline
  Dell-XPS L421X& \pbox{50cm}{Ubuntu \\ (V.14.04 LTS) }&   \pbox{50cm}{1.8GHz*4 \\ Intel Core  i5} & 8 GB \\ \hline
Dell-Inspiron-3721   & \pbox{50cm}{Ubuntu \\ (V.14.04 LTS)} &   \pbox{50cm}{2.00GHz*4 \\Intel Core  i5 }& 4 GB      \\\hline
 Dell-Inspiron-3521 & \pbox{50cm}{ Ubuntu \\(V.16.04 LTS)}    &\pbox{50cm}{ 1.8 GHz*4 \\Intel Core i5 } &  6 GB \\\hline
    iMac-Early 2010 & \pbox{50cm}{cinux Mint \\ (V.17.1 Rebecca) }&   \pbox{50cm}{3.06GHz*2} & 4 GB   \\ \hline  
Dell-Inspiron-5559   & \pbox{50cm}{Ubuntu \\ (V.16.04 LTS) }&   \pbox{50cm}{2.30GHz*4 \\ Intel Core i5}  & 8 GB  \\\hline
iMac-Early 2009 & \pbox{50cm}{OS X El Capitan \\(V.10.11.6)}    &\pbox{50cm}{ 2.93 *2 GHz \\Intel Core Due}&  8 GB     \\\hline
MacBook Air     & \pbox{50cm}{OS X El Capitan \\(V.10.11.3)}    &\pbox{50cm}{ 1.6 *2 GHz \\Intel Core i5 }&  8 GB     \\\hline
\end{tabular}
\end{table}

We used two benchmarks of datasets  from Chameleon ~\cite{Pasi-15}. These are commonly used
to test and evaluate clustering. Table \ref{DS} gives details about the datasets.

\begin{table}[!htb]
\centering
\caption{Datasets}
\label{DS}

\begin{tabular}{|c|c|c|}
\hline
 \textbf{Benchmark}& \textbf{Size} & \textbf{Descriptions} \\\hline
 \textbf{D1}& 10,000 Points & \begin{tabular}[c]{@{}c@{}}Different shapes, with \\ some clusters surrounded\\  by others\end{tabular} \\\hline
 \textbf{D2}& 30,000 Points& \begin{tabular}[c]{@{}c@{}}2 small circles,\\ 1 big circle\\ and 2 linked ovals\end{tabular} \\ \hline
\end{tabular}
\end{table}

The DDC approach is tested using  various partitions of different sizes. Various scenarios
were created based on  the goals of the experiments. These scenarios  mainly differ on the
way the datasets are  divided among the processing nodes of  the distributed platform. For
each scenario, we recorded  the execution time for the local  clustering, the merging step
including contour calculations, aggregation time and  Idle time.). finally we capture also
the total execution time that the approach takes to finish all the steps. in the following
we describe the different scenarios considered.

 \subsection{Experiment I}
 
 In this  scenario we give  each machine a  random chunk of the  dataset, the size  of the
 partition that was generated  for each machine is in the range  between $1500$ points and
 $10000$  points. As  the dataset  is relatively  small we  chose eight  machines for  the
 computing platform.
 
Table \ref{T1} shows the  execution time taken by each machine to  run the algorithm (step
one and step two) using synchronous  and asynchronous communications respectively, it also
shows the overall time taken to finish all the steps.  

From Table  \ref{T1}, we can see  that the time taken  by each machine to  accomplish the
first step of the algorithm is the same for both synchronous and asynchronous, whereas the
time  of the  second step  is different.   We can  also notice  that each  machine returns
different  execution time  of  the whole  algorithm.  This is  because  the machines  have
different capacities (see Table \ref{capa}).

The total execution time of the algorithm while using synchronous communication is smaller
compared  to  when  using  synchronous  communication. This  is  because  in  synchronous
communications, machines have more waiting time (up to 60\% waiting time).

\begin{table*}[!htb]
\centering
\caption{Time  (ms) taken  by  eight machines  to  run scenario  I  using synchronous  and
  asynchronous communications} 
\label{T1}
\begin{tabular}{cccccccc}
\cline{3-8}
  & \multicolumn{1}{c|}{}     & \multicolumn{3}{c||}{\textbf{Synchronous}}                                                            & \multicolumn{3}{c|}{\textbf{Asynchronous}}                                                           \\ \hline
\multicolumn{1}{|c|}{\textbf{Machine}} & \multicolumn{1}{c|}{\textbf{DS Size}} & \multicolumn{1}{c|}{\textbf{STEP01}} & \multicolumn{1}{c|}{\textbf{STEP2}} & \multicolumn{1}{c||}{\textbf{Time}}  & \multicolumn{1}{c|}{\textbf{STEP1}} & \multicolumn{1}{c|}{\textbf{STEP2}} & \multicolumn{1}{c|}{\textbf{Time}}  \\ \hline
\multicolumn{1}{|c|}{\textbf{M1}}      & \multicolumn{1}{c|}{10000}   & \multicolumn{1}{c|}{21270} & \multicolumn{1}{c|}{1104}  & \multicolumn{1}{c||}{22374} & \multicolumn{1}{c|}{21270} & \multicolumn{1}{c|}{554}    & \multicolumn{1}{c|}{21824} \\ \hline
\multicolumn{1}{|c|}{\textbf{M2}}      & \multicolumn{1}{c|}{2500}    & \multicolumn{1}{c|}{1060}  & \multicolumn{1}{c|}{20862} & \multicolumn{1}{c||}{21922} & \multicolumn{1}{c|}{1060}  & \multicolumn{1}{c|}{2515}  & \multicolumn{1}{c|}{3575}  \\ \hline
\multicolumn{1}{|c|}{\textbf{M3}}      & \multicolumn{1}{c|}{3275}    & \multicolumn{1}{c|}{5093}  & \multicolumn{1}{c|}{16930} & \multicolumn{1}{c||}{22023} & \multicolumn{1}{c|}{5093}  & \multicolumn{1}{c|}{2017}  & \multicolumn{1}{c|}{7110}  \\ \hline
\multicolumn{1}{|c|}{\textbf{M4}}      & \multicolumn{1}{c|}{5000}    & \multicolumn{1}{c|}{4592}  & \multicolumn{1}{c|}{17644} & \multicolumn{1}{c||}{22236} & \multicolumn{1}{c|}{4591}  & \multicolumn{1}{c|}{2620}  & \multicolumn{1}{c|}{7211}  \\ \hline
\multicolumn{1}{|c|}{\textbf{M5}}      & \multicolumn{1}{c|}{1666}    & \multicolumn{1}{c|}{227}   & \multicolumn{1}{c|}{21642} & \multicolumn{1}{c||}{21869} & \multicolumn{1}{c|}{227}   & \multicolumn{1}{c|}{391}   & \multicolumn{1}{c|}{618}   \\ \hline
\multicolumn{1}{|c|}{\textbf{M6}}      & \multicolumn{1}{c|}{2000}    & \multicolumn{1}{c|}{292}   & \multicolumn{1}{c|}{21736} & \multicolumn{1}{c||}{22028} & \multicolumn{1}{c|}{292}   & \multicolumn{1}{c|}{416}   & \multicolumn{1}{c|}{708}   \\ \hline
\multicolumn{1}{|c|}{\textbf{M7}}      & \multicolumn{1}{c|}{5000}    & \multicolumn{1}{c|}{7520}  & \multicolumn{1}{c|}{14665} & \multicolumn{1}{c||}{22185} & \multicolumn{1}{c|}{7515}  & \multicolumn{1}{c|}{13949} & \multicolumn{1}{c|}{21464} \\ \hline
\multicolumn{1}{|c|}{\textbf{M8}}      & \multicolumn{1}{c|}{1500}    & \multicolumn{1}{c|}{200}   & \multicolumn{1}{c|}{21842} & \multicolumn{1}{c||}{22042} & \multicolumn{1}{c|}{195}   & \multicolumn{1}{c|}{4605}  & \multicolumn{1}{c|}{4800}  \\ \hline
\hline
 \cline{3-8} 
\multicolumn{1}{r}{}          & \multicolumn{1}{c|}{}        & \multicolumn{2}{c|}{\textbf{Total Exec-Time}}                     & \multicolumn{1}{c||}{\textbf{22374}} & \multicolumn{2}{r|}{\textbf{Total Exec-Time}}                     & \multicolumn{1}{c|}{\textbf{21824}} \\ \cline{3-8} 
\end{tabular}
\end{table*}

 \subsection{Experiment II}

 In this  scenario we allocate  the whole  dataset size to  one machine and  the remaining
 machines were allocated  one eight of the  dataset each. This scenario is  chosen to show
 the worst case of waiting time.
 
Table \ref{T2} shows the execution time taken by each machine to execute the DDC technique
(step one and step two) using synchronous and asynchronous communications respectively, it
also shows the overall time taken to finish all the steps.  

From Table \ref{T2}, we can notice that  the difference between the execution times of the
synchronous  and  asynchronous  DDC  is   still  significant.   Because  with  synchronous
communications the  machines need to wait  for the last  machine to finish its  first step
before they all start  merging their results (step 2), whereas  for asynchronous model the
seven machines  did the  merging (step2)  while the last  machine finishes  its clustering
(step1).

%....................................S2-T1...................................................

\begin{table}[!htb]
\centering
\caption{Time (ms) taken by eight machines to run scenario II using synchronous and asynchronous communications}
\label{T2}
\begin{tabular}{cc|c|c|c||c|c|c|}
\cline{3-8}
\multicolumn{1}{c}{}          &                                       & \multicolumn{3}{c||}{\textbf{Synchronous}}                                                                      & \multicolumn{3}{c|}{\textbf{Asynchronous}}                                                                     \\ \hline
\multicolumn{1}{|c|}{Machine} & \multicolumn{1}{c|}{\textbf{DS Size}} & \multicolumn{1}{c|}{\textbf{STEP1}} & \multicolumn{1}{c|}{\textbf{STEP2}} & \multicolumn{1}{c||}{\textbf{Time}} & \multicolumn{1}{c|}{\textbf{STEP1}} & \multicolumn{1}{c|}{\textbf{STEP2}} & \multicolumn{1}{c|}{\textbf{Time}} \\ \hline
\multicolumn{1}{|c|}{M1}      & 10000                                 & 21270                               & 973                                 & 22243                              & 21270                               & 595                                 & 21865                              \\ \hline
\multicolumn{1}{|c|}{M2}      & 1250                                  & 215                                 & 21775                               & 21990                              & 215                                 & 518                                 & 733                                \\ \hline
\multicolumn{1}{|c|}{M3}      & 1250                                  & 640                                 & 21383                               & 22023                              & 640                                 & 20100                               & 20740                              \\ \hline
\multicolumn{1}{|c|}{M4}      & 1250                                  & 304                                 & 21730                               & 22034                              & 304                                 & 497                                 & 801                                \\ \hline
\multicolumn{1}{|c|}{M5}      & 1250                                  & 161                                 & 22034                               & 22195                              & 161                                 & 394                                 & 555                                \\ \hline
\multicolumn{1}{|c|}{M6}      & 1250                                  & 171                                 & 21856                               & 22027                              & 170                                 & 286                                 & 456                                \\ \hline
\multicolumn{1}{|c|}{M7}      & 1250                                  & 245                                 & 21918                               & 22163                              & 245                                 & 509                                 & 754                                \\ \hline
\multicolumn{1}{|c|}{M8}      & 1250                                  & 185                                 & 21854                               & 22039                              & 185                                 & 858                                 & 1043                               \\ \hline \hline
\multicolumn{1}{c}{}          & \textbf{}                             & \multicolumn{2}{c|}{\textbf{Total Exec-Time}}                             & \textbf{22243}                     & \multicolumn{2}{c|}{\textbf{Total Exec-Time}}                             & \textbf{21865}                     \\ \cline{3-8} 
\end{tabular}
\end{table}

 \subsection{Experiment III}
 
 In this scenario we allocate to seven machines  the whole dataset and the one machine was
 allocated one  eight of the dataset.  This scenario is chosen  to show the effect  of the
 complexity of the local clustering complexity on  the machines and on the waiting time of
 some powerful machines.
 
Table \ref{T3} shows the  execution time taken by each machine to  run the algorithm (step
one and step two) using synchronous  and asynchronous communications respectively, it also
shows the overall time taken to finish all the steps.  

This scenario  is the opposite  of the previous  scenario. Unlike the  previous scenarios,
Table \ref{T3}  shows that the difference  between the execution times  of synchronous and
asynchronous versions of the  DDC is smaller.  This is because in  both cases the machines
spend  more time  finishing  the first  step,  therefore,  the waiting  time  is less  for
synchronous over asynchronous model.

%....................................S3-T1...................................................
 \begin{table}[!htb]
\centering
\caption{Time (ms) taken by eight machines to run scenario III using synchronous and asynchronous communications}
\label{T3}
\begin{tabular}{cccccccc}
\cline{3-8}
\multicolumn{2}{c|}{}                                                      & \multicolumn{3}{c||}{\textbf{Synchronous}}                                                                       & \multicolumn{3}{c|}{\textbf{Asynchronous}}                                                                      \\ \hline
\multicolumn{1}{|c|}{\textbf{Machine}}     & \multicolumn{1}{c|}{\textbf{DS Size}} & \multicolumn{1}{c|}{\textbf{STEP1}} & \multicolumn{1}{c|}{\textbf{STEP2}} & \multicolumn{1}{c||}{\textbf{Time}}  & \multicolumn{1}{c|}{\textbf{STEP1}} & \multicolumn{1}{c|}{\textbf{STEP2}} & \multicolumn{1}{c|}{\textbf{Time}}  \\ \hline
\multicolumn{1}{|c|}{\textbf{M1}} & \multicolumn{1}{c|}{10000}            & \multicolumn{1}{c|}{21270}          & \multicolumn{1}{c|}{35978}          & \multicolumn{1}{c||}{57248}          & \multicolumn{1}{c|}{21270}          & \multicolumn{1}{c|}{905}            & \multicolumn{1}{c|}{22175}          \\ \hline
\multicolumn{1}{|c|}{\textbf{M2}} & \multicolumn{1}{c|}{10000}            & \multicolumn{1}{c|}{21590}          & \multicolumn{1}{c|}{34869}          & \multicolumn{1}{c||}{56459}          & \multicolumn{1}{c|}{21590}          & \multicolumn{1}{c|}{11513}          & \multicolumn{1}{c|}{33103}          \\ \hline
\multicolumn{1}{|c|}{\textbf{M3}} & \multicolumn{1}{c|}{10000}            & \multicolumn{1}{c|}{53005}          & \multicolumn{1}{c|}{3008}           & \multicolumn{1}{c||}{56013}          & \multicolumn{1}{c|}{53005}          & \multicolumn{1}{c|}{3292}           & \multicolumn{1}{c|}{56297}          \\ \hline
\multicolumn{1}{|c|}{\textbf{M4}} & \multicolumn{1}{c|}{10000}            & \multicolumn{1}{c|}{32424}          & \multicolumn{1}{c|}{24691}          & \multicolumn{1}{c||}{57115}          & \multicolumn{1}{c|}{32424}          & \multicolumn{1}{c|}{6996}           & \multicolumn{1}{c|}{39420}          \\ \hline
\multicolumn{1}{|c|}{\textbf{M5}} & \multicolumn{1}{c|}{10000}            & \multicolumn{1}{c|}{17364}          & \multicolumn{1}{c|}{38493}          & \multicolumn{1}{c||}{55857}          & \multicolumn{1}{c|}{17364}          & \multicolumn{1}{c|}{4612}           & \multicolumn{1}{c|}{21976}          \\ \hline
\multicolumn{1}{|c|}{\textbf{M6}} & \multicolumn{1}{c|}{10000}            & \multicolumn{1}{c|}{15841}          & \multicolumn{1}{c|}{41237}          & \multicolumn{1}{c||}{57078}          & \multicolumn{1}{c|}{15841}          & \multicolumn{1}{c|}{2066}           & \multicolumn{1}{c|}{17907}          \\ \hline
\multicolumn{1}{|c|}{\textbf{M7}} & \multicolumn{1}{c|}{10000}            & \multicolumn{1}{c|}{38732}          & \multicolumn{1}{c|}{18483}          & \multicolumn{1}{c||}{57215}          & \multicolumn{1}{c|}{38727}          & \multicolumn{1}{c|}{18459}          & \multicolumn{1}{c|}{57186}          \\ \hline
\multicolumn{1}{|c|}{\textbf{M8}} & \multicolumn{1}{c|}{1250}             & \multicolumn{1}{c|}{185}            & \multicolumn{1}{c|}{56915}          & \multicolumn{1}{c||}{57100}          & \multicolumn{1}{c|}{184}            & \multicolumn{1}{c|}{16077}          & \multicolumn{1}{c|}{16261}          \\ \hline \hline
\multicolumn{1}{c}{}              & \multicolumn{1}{c|}{}                 & \multicolumn{2}{c|}{\textbf{Total Exec-Time}}                             & \multicolumn{1}{c||}{\textbf{57248}} & \multicolumn{2}{c|}{\textbf{Total Exec-Time}}                             & \multicolumn{1}{c|}{\textbf{57186}} \\ \cline{3-8}                                 
\end{tabular}
\end{table}

\subsection{Experiment IV}
 
In this scenario we took into account the machines capabilities and we divide the datasets
according to their  capacities.  Therefore the work load is  evenly distributed among them
and we expect them to finish the first phase more or less at the same time. This allows to
reduce the waiting time of the machines  and follow immediately with the second phase. The
total execution times  of synchronous and asynchronous versions should  be the same.  This
case favours more the synchronous implementation of the approach.
 
As predicted, Table \ref{T4} shows that there is no significant difference between the two
execution times. Note that  the little difference in favour of  the synchronous version is
due to  the fact that  in the asynchronous  model the machines  still need to  execute the
algorithm that checks which one finished first and receive the contours  for merging.

%....................................S4-T1...................................................
\begin{table}[!htb]
\centering
\caption{Time (ms) taken by eight machines to run scenario IV using synchronous and asynchronous communications}
\label{T4}
\begin{tabular}{ccccccccc}
\cline{3-8}
\multicolumn{2}{c|}{}                                                      & \multicolumn{3}{c||}{\textbf{Synchronous}}                                                                      & \multicolumn{3}{c|}{\textbf{Asynchronous}}                                                                              \\ \hline
\multicolumn{1}{|c|}{\textbf{Machine}}     & \multicolumn{1}{c|}{\textbf{DS Size}} & \multicolumn{1}{c|}{\textbf{STEP1}} & \multicolumn{1}{c|}{\textbf{STEP2}} & \multicolumn{1}{c||}{\textbf{Time}} & \multicolumn{1}{c|}{\textbf{STEP1}} & \multicolumn{1}{c|}{\textbf{STEP2}} & \multicolumn{1}{c|}{\textbf{Time}} \\ \hline
\multicolumn{1}{|c|}{\textbf{M1}} & \multicolumn{1}{c|}{1500}             & \multicolumn{1}{c|}{256}            & \multicolumn{1}{c|}{1505}           & \multicolumn{1}{c||}{1761}          & \multicolumn{1}{c|}{256}            & \multicolumn{1}{c|}{1159}           & \multicolumn{1}{c|}{1415}          \\ \hline
\multicolumn{1}{|c|}{\textbf{M2}} & \multicolumn{1}{c|}{1660}             & \multicolumn{1}{c|}{260}            & \multicolumn{1}{c|}{598}            & \multicolumn{1}{c||}{858}           & \multicolumn{1}{c|}{260}            & \multicolumn{1}{c|}{1512}           & \multicolumn{1}{c|}{1772}          \\ \hline
\multicolumn{1}{|c|}{\textbf{M3}} & \multicolumn{1}{c|}{500}              & \multicolumn{1}{c|}{252}            & \multicolumn{1}{c|}{1061}           & \multicolumn{1}{c||}{1313}          & \multicolumn{1}{c|}{252}            & \multicolumn{1}{c|}{626}            & \multicolumn{1}{c|}{878}           \\ \hline
\multicolumn{1}{|c|}{\textbf{M4}} & \multicolumn{1}{c|}{1000}             & \multicolumn{1}{c|}{253}            & \multicolumn{1}{c|}{621}            & \multicolumn{1}{c||}{874}           & \multicolumn{1}{c|}{253}            & \multicolumn{1}{c|}{608}            & \multicolumn{1}{c|}{861}           \\ \hline
\multicolumn{1}{|c|}{\textbf{M5}} & \multicolumn{1}{c|}{1500}             & \multicolumn{1}{c|}{255}            & \multicolumn{1}{c|}{1492}           & \multicolumn{1}{c||}{1747}          & \multicolumn{1}{c|}{255}            & \multicolumn{1}{c|}{600}            & \multicolumn{1}{c|}{855}           \\ \hline
\multicolumn{1}{|c|}{\textbf{M6}} & \multicolumn{1}{c|}{1400}             & \multicolumn{1}{c|}{260}            & \multicolumn{1}{c|}{605}            & \multicolumn{1}{c||}{865}           & \multicolumn{1}{c|}{260}            & \multicolumn{1}{c|}{514}            & \multicolumn{1}{c|}{774}           \\ \hline
\multicolumn{1}{|c|}{\textbf{M7}} & \multicolumn{1}{c|}{1000}             & \multicolumn{1}{c|}{259}            & \multicolumn{1}{c|}{1030}           & \multicolumn{1}{c||}{1289}          & \multicolumn{1}{c|}{259}            & \multicolumn{1}{c|}{939}            & \multicolumn{1}{c|}{1198}          \\ \hline
\multicolumn{1}{|c|}{\textbf{M8}} & \multicolumn{1}{c|}{1500}             & \multicolumn{1}{c|}{250}            & \multicolumn{1}{c|}{603}            & \multicolumn{1}{c||}{853}           & \multicolumn{1}{c|}{250}            & \multicolumn{1}{c|}{1500}           & \multicolumn{1}{c|}{1750}          \\ \hline \hline
\multicolumn{1}{c}{}              & \multicolumn{1}{c|}{}                 & \multicolumn{2}{c|}{\textbf{Total Exec Time}}                             & \multicolumn{1}{c||}{\textbf{1761}} & \multicolumn{2}{c|}{\textbf{Total Exec Time}}                             & \multicolumn{1}{c|}{\textbf{1772}} \\ \cline{3-8} 
\end{tabular}
\end{table}

\subsection{ Effective Speedup}
The goal here is  to compare our parallel clustering to the  sequential algorithm and show
the  DDC speedup  over the  sequential  version of  clustering, as  mentioned in  Equation
\ref{eq:3}.

Considering the best scenario of executing the sequential version of DBSCAN on the fastest
machine in the system. For instance, $T_1=  15841$ ms.  Clustering a partition of the same
dataset on the same machine  will take $ = T^d_1= 258$ ms.  The  execution time of the DDC
on the same datasets on eight heterogeneous machines with load balancing is $T_p= 1761$ ms
(see Table  \ref{T4}).  Therefore, from  equation \ref{eq:3}, we  can deduce a  speedup of
$9$, which  is still a  super-linear speedup. In  the next section  we will show  how many
processing nodes are required to cluster a dataset of size $N$.

\subsection{Scalability}
\label{sec:Scal}

\begin{figure}[!htb]
\begin{center}
\includegraphics[width=.8\textwidth]{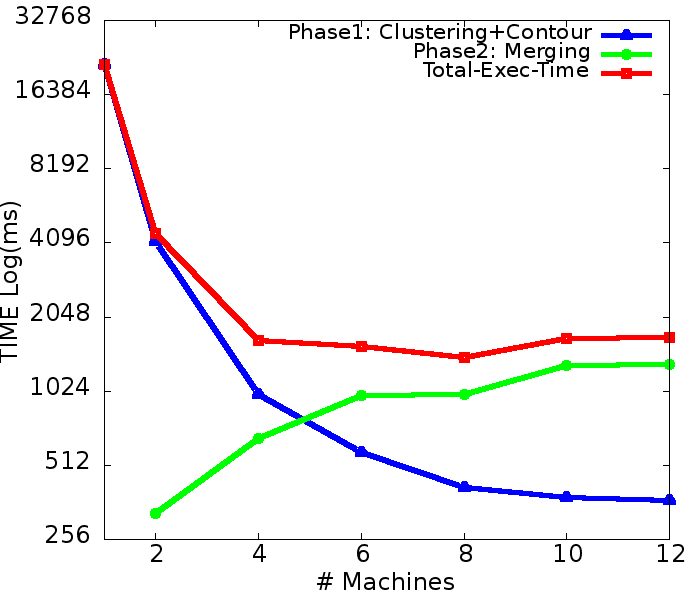}
\caption{Scalability Experiment using dataset $T_1$.}
\label{SCAL1}
\end{center}
\end{figure}

\begin{figure}[!htb]
\begin{center}
\includegraphics[width=.8\textwidth]{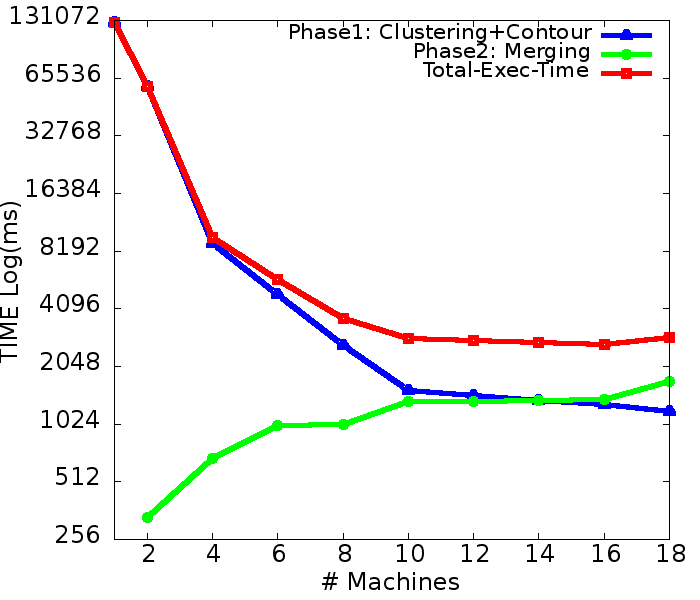}
\caption{Scalability Experiment using dataset $T_2$.}
\label{SCAL2}
\end{center}
\end{figure}

The goal here  is to show that the  DDC technique scales well and also  we can dynamically
determine the  optimal number of  processing nodes required to  cluster a dataset  of size
$N$. We consider two  datasets, the first dataset $D_1$ contains  $10,000$ data points and
the second  $D_1$ contains $30,000$ data  points.  Figure \ref{SCAL1} shows  the execution
time ($y\_axis$  is in $log_2$)  against the  number of machines  in the system  using the
first dataset  and Figure \ref{SCAL2} shows  the execution time ($y\_axis$  is in $log_2$)
against the number  of machines in the  system using the second  dataset contains $30,000$
data points.

%..............................................................................

As one can see,  from both Figures \ref{SCAL1} and \ref{SCAL2}, the  execution time of the
first phase  (Clustering and Contour)  keeps decreasing as the  number of machines  in the
distributed  system increases.   However, the  time of  the second  phase (merging)  keeps
increasing gradually with the number of machines in the distributed system that is because
the amount  of communications in  the second phase increases  when the number  of machines
increases.

In addition, the total execution time of the algorithm (which is the sum of the two times,
phase one and  two) keep decreasing as  the number of processing nodes  increases until it
reaches a certain points where the total  execution time starts to increase (at 8 machines
for dataset $D_1$ and at 16 machines  for dataset $D_2$). The optimal number of processing
nodes required to  execute DDC is returned  when the overhead of the  approach exceeds the
execution time of the local clustering. This  is a very interesting characteristic, as one
can determine the number of machines that can be allocated in advance.

\section{Conclusion}
\label{sec:Con}

In this paper, we proposed an efficient and flexible distributed clustering framework that
can work with existing data mining  algorithms. The approach exploits the processing power
of  the   distributed  platform   by  maximising  the   parallelism  and   minimising  the
communications and mainly the size of the data  that is exchanged between the nodes of the
system. It is implemented using both  synchronous and asynchronous communications, and the
results  were significantly  in favour  of  the asynchronous  model. The  approach has  an
efficient data reduction phase which reduces  significantly the size of the data exchanged
therefore, it deals  with the problem of  communication overhead.  The DDC  approach has a
super-linear speedup when the complexity of the  local clustering has an NP complexity. We
also can  determine the optimal number  of processing nodes in advance.

\section*{Acknowledgement}
\label{sec:Ack}
The  research work  is  conducted in  the  Insight  Centre for  Data  Analytics, which  is
supported by Science Foundation Ireland under Grant Number SFI/12/RC/2289.

\bibliographystyle{splncs03}
\bibliography{AUSDM-References}

\end{document}